\documentclass[
 preprint,nofootinbib,
 prd,
]{revtex4-1}

\usepackage{graphicx}% Include figure files
\usepackage{bm}% bold math
\usepackage{color}
\usepackage{hyperref}% add hypertext capabilities
\usepackage{amsmath}
\usepackage{bm}
\usepackage{slashed}
\usepackage{ulem}% for \sout{}
\usepackage[utf8]{inputenc}

\newcommand{\beq}{\begin{equation}}
\newcommand{\eeq}{\end{equation}}

\newcommand{\mt}{\tilde\mu}

\newcommand{\sgn}{\mathrm{sgn}}

\newcommand{\tw}{\textwidth}
\newcommand{\fig}[1]{Fig.~\ref{#1}}

\newcommand{\Eq}[1]{Eq.~\eqref{#1}}
\newcommand{\ave}[1]{\langle{#1}\rangle}
\renewcommand{\sec}[1]{Sec.~\ref{#1}}

\def\x{{\bf x}}

\def\e{{\bf e}}
\def\tmu{\tilde{\mu}}

\def\eg{{e.g.}}
\def\cfr{{cf.}}

\begin{document}

\title{Influence of vector interactions on the
favored shape of inhomogeneous chiral condensates}
% Force line breaks with \\
% \thanks{A footnote to the article title}%

\author{Stefano Carignano}
\affiliation{%
Instituto de Ciencias del Espacio (ICE, CSIC)
C. Can Magrans s.n., 08193 Cerdanyola del Vall\`es, Catalonia, Spain
}%
\author{Marco Schramm}
%  \altaffiliation{Institut f\"ur Kernphysik, Technische Universit\"at Darmstadt, Germany}%Lines break automatically or can be forced with \\
\author{Michael Buballa}%
%  \email{Second.Author@institution.edu}
\affiliation{%
Theoriezentrum, Institut f\"ur Kernphysik, Technische Universit\"at Darmstadt, Germany
}%

\date{\today}% It is always \today, today,
             %  but any date may be explicitly specified

\begin{abstract}

We update a previous study of the effects of vector interactions in the
Nambu--Jona-Lasinio model on the formation of
inhomogeneous chiral symmetry breaking condensates. In particular, by
properly considering a spatially modulated vector
mean-field associated with the quark number density of the system we show
that, as the value of the vector coupling increases, a chiral density wave
modulation can become thermodynamically favored over a real sinusoidal
modulation. This behavior is found both via a Ginzburg-Landau analysis
close to the Lifshitz point, as well as with a full numerical
diagonalization of the mean-field Dirac Hamiltonian at vanishing
temperature.

\end{abstract}

\maketitle

\section{Introduction}

The phase structure of QCD at high densities and intermediate temperatures is expected to be extremely rich.
A particularly fascinating scenario suggested by explicit model calculations revolves around the formation of 
an inhomogeneous ``island'', characterized by the presence of crystalline chiral condensates, 
which
 can appear in the region where  the first-order chiral phase transition is expected to take place (for a recent review, see \cite{Buballa:2014tba}). 
As a consequence, the critical point (CP) as a cornerstone of the phase diagram is replaced by a Lifshitz point (LP) where three different phases - a homogeneous broken, an inhomogeneous and a chirally restored one - coexist.  
Among various shapes which have been compared so far, the energetically favored crystalline structure within the inhomogeneous island is the so-called ``real-kink crystal'' (RKC) \cite{Nickel:2009wj,Carignano:2012sx}, which assumes the form of an array of domain-wall solitons close to the phase transition to the homogeneous phase and rapidly evolves into a sinusoidal shape as one approaches chiral restoration.
The existence of an inhomogeneous phase might have phenomenological consequences for compact stars \cite{Buballa:2015awa}
and for the physics of heavy-ion collisions, especially for the future facilities at FAIR and NICA as well as the beam-energy scan at RHIC~\cite{Pisarski:2018bct,Carlomagno:2018ogx}.

A common extension considered in effective models of QCD such as the Nambu--Jona-Lasinio (NJL) model for describing matter at high densities is the inclusion of a repulsive vector interaction channel \cite{Asakawa:1989,Buballa:1996tm,Fukushima:2008,Fukushima:2008b,Zhang:2009mk}.  Within the mean-field approximation, the time component of the vector condensate can be identified with the density of the system, and induces an effective shift in the chemical potential of the quarks \cite{Asakawa:1989}.
 If only homogeneous matter is considered, aside from moving the chiral transition line to higher chemical potentials, the most notable effect of vector interactions is a shift of the critical point towards lower temperatures, leading to its disappearance beyond a certain value of the vector coupling constant \cite{Fukushima:2008b}. 

A first investigation of the effects of vector interactions on chiral crystalline phases has been performed in Ref.~\cite{CNB:2010}. There it was found that the inclusion of vector interactions leads to a dramatic enhancement of the inhomogeneous region in the model phase diagram
and the disappearance of the critical point inside the inhomogeneous phase.
In that study, in order to be able to use known analytical expressions for the eigenvalue spectrum of the Dirac Hamiltonian in the presence of certain crystalline backgrounds, instead of considering a spatially modulated density, the vector mean field was approximated by its spatial average. While this approach is still self-consistent when a plane-wave modulation of the chiral condensate (typically referred to as a ``(dual) chiral density wave" (CDW) \cite{NT:2004}) is considered, this is certainly not the case for modulations like the RKC \cite{CNB:2010} or even a simple sinusoidal ansatz.  
For these cases, a repulsive vector interaction might generate an additional energy cost for the formation of a spatially modulated quark density, possibly influencing the competition between different crystalline phases and altering the resulting phase structure of the model.
In this work, we aim at investigating this possibility by lifting the assumption of constant density and allow for a more consistent treatment of spatially modulated vector condensates within the model. 

This paper is structured as follows:
 We begin by describing our model and the procedure for including inhomogeneous chiral and vector condensates in \sec{sec:theory}, while in
\sec{sec:gl} and \sec{sec:num} we present results from a Ginzburg-Landau investigation and from a numerical diagonalization of the Dirac Hamiltonian, respectively. We finally conclude in \sec{sec:conclusions}.

\section{Model and basic formalism \label{sec:theory}}

We consider an NJL model in the chiral limit, given by the Lagrangian 
\begin{align}
 \mathcal L=\bar\psi i\slashed\partial \psi + G_S\left(\left(\bar\psi\psi\right)^2+\left(\bar\psi i \gamma^5\tau^a\psi\right)^2\right) - G_V\left(\bar\psi\gamma^\mu\psi\right)^2\,,
 \end{align}
where $\psi$ denotes the quark field with vanishing bare mass,
$N_f=2$ flavor and $N_c=3$ color degrees of freedom.
In addition to the standard isoscalar scalar and isovector pseudoscalar interactions with coupling constant $G_S$,
the Lagrangian contains an interaction in the isoscalar vector channel with coupling constant $G_V$.
Here  $\gamma^\mu$ are Dirac matrices, and $\tau^a$ are Pauli matrices in flavor space.

 Performing the mean-field approximation, we allow for 
 the scalar $\ave {\bar\psi(\x)\psi(\x)} \equiv S(\x)$, pseudoscalar  $\ave {\bar\psi(\x) i\gamma^5\tau^a\psi(\x)} \equiv P(\x) \delta_{a3}$ and vector $\ave {\bar\psi(\x) \gamma^\mu\psi(\x)} \equiv n(\x) g^{\mu0}$ condensates, which are restricted to be time-independent but allowed to vary in space
 in order to describe possible inhomogeneities.
As in Ref.~\cite{CNB:2010}, we assume that the pseudoscalar condensate is diagonal in flavor space and 
consider only the $0$-component  of the vector condensate, 
corresponding to the quark-number density $n(\x) =  \ave {\bar\psi(\x) \gamma^0\psi(\x)}$.
This latter assumption is well justified when considering spatially uniform matter since non-vanishing spatial vector components 
break the rotational invariance of the system.
While this argument is no longer valid in phases where the rotational symmetry is already broken 
by the scalar condensates, vanishing space-like components of the vector condensate could also be protected 
if the ground state is invariant under parity transformations. 
This is the case for all modulations considered in this article. Indeed preliminary studies seem to confirm that 
 $\ave{\bar\psi \gamma^k \psi} = 0$,   $k=1,2,3$, holds for these phases~\cite{Marco:MSc,Tobias:MSc}.

Following standard procedures (see Ref.~\cite{Buballa:2014tba} for details),
the mean-field thermodynamic potential per unit volume $V$ at a given temperature $T$ and chemical potential $\mu$ 
can then be written as a sum of a kinetic and a  condensate  contribution, 
\beq 
 \Omega(T,\mu)=\Omega_{kin}(T,\mu) + \Omega_{cond} \,,
 \eeq 
where the latter is simply given by a volume integral over the condensates, 
\beq
 \Omega_{cond}
 =\frac{1}{V}\int_V d^3 x\big[ G_S  \left(S^2(\x) + P^2(\x)\right) - G_V n^2(\x)\big] \,.
\label{eq:Omegacond}
\eeq
The kinetic contribution,
\beq
  \Omega_{kin}
 =-\frac{1}{V}\sum_{E_\lambda} T\ln\left[2\cosh\left(\frac{E_\lambda-\mu}{2T}\right)\right] \,,     
 \label{eq:Omegakin}
\eeq
basically amounts to a sum over the eigenvalues $E_\lambda$ of the mean-field Dirac Hamiltonian
\beq
        H(\x) = H_0(\x) + 2G_V n(\x)\,,
\label{eq:H}
\eeq
where 
\beq
       H_0(\x) = \gamma^0 \left[ -i{\bm \gamma\cdot\partial}  + P_+M(\x) + P_- M^*(\x) \right]
\eeq
corresponds to the mean-field Hamiltonian without vector interactions. 
Here 
\beq
  M(\x)=  2G_S\left(S(\x)+iP(\x)\right) \,,\label{eq:Mz}
\eeq
is a complex constituent quark mass function and 
$P_\pm = \frac{1}{2}\left( 1 \pm \gamma^5 \tau^3 \right)$
are chirality projectors.
Generalizing a method often employed in homogeneous phases~\cite{Asakawa:1989},
we can also absorb the effects of the vector mean field in \Eq{eq:H} into a  space-dependent shifted chemical potential function,
\beq
 \tmu(\x) =\mu-2G_V n(\x) \,,
\label{eq:tmudef}
\eeq
so that
\beq
       H(\x) - \mu = H_0(\x) - \tmu(\x)\,.
       \label{eq:Hofx}
\eeq
The condensate part of the thermodynamic potential, \Eq{eq:Omegacond}, is then given by
\beq
 \Omega_{cond}
 =\frac{1}{V}\int_V d^3 x\left(\frac{|M(\x)|^2}{4G_S} - \frac{(\tmu(\x)-\mu)^2}{4G_V}\right) \,.
\eeq

Already for the simpler case without vector interactions,
the diagonalization of the Dirac Hamiltonian in the presence of inhomogeneous condensates is a non-trivial task, 
as an explicit space dependence of the mean fields leads to a non-diagonal structure in momentum space.  
For two particular one-dimensional spatial modulations of the order parameters, however, it is possible to obtain an analytical expression for the eigenvalue spectrum of the Hamiltonian when vector interactions are switched off. 
The first one  
is the CDW \cite{NT:2004}, which is a simple plane-wave modulation
 \begin{align}
 M_{CDW}(z)= %M_1
 \frac{\Delta}{\sqrt{2}} \exp(i q z) \,,
 \label{eq:mzCDW}
\end{align}
where, without loss of generality, we have chosen the wave to vary along the $z$-direction. 
The amplitude $\Delta/\sqrt{2}$
and the wave number $q$ 
have to be detemined as functions of $T$ and $\mu$ by minimizing 
the thermodynamic potential. 

The other example is {the} one-dimensional {RKC}, which can be expressed in terms of Jacobi elliptic functions,
\begin{align}
 M_{RKC}(z)=\Delta\nu \, \frac{\mathrm{sn}(\Delta z|\nu)\mathrm{cn}(\Delta z|\nu)}{\mathrm{dn}(\Delta z|\nu)} \,,
 \label{eq:mzsoli}
\end{align}
with parameters $\Delta$ and $\nu$, which are again determined by minimizing $\Omega$.
The so-called elliptic modulus $\nu$ thereby determines the shape of the mass function.  
It is found that
this type of solution assumes a solitonic shape close to the onset of the inhomogeneous phase at low chemical potentials, 
while as the chemical potential increases it quickly assumes a sinusoidal shape, see \cite{CNB:2010,Carignano:2017uyp}. 
 
In the absence of vector interactions, the RKC solution is energetically favored over the CDW throughout the 
entire inhomogeneous window \cite{Nickel:2009wj}. 
It is also more favored than all other shapes (including higher-dimensional modulations) which have been tested so far either
by numerical diagonalization of the Hamiltonian \cite{Carignano:2012sx} or within Ginzburg-Landau studies \cite{Abuki:2011pf,Carignano:2017meb}.
%%%

In general, the inclusion of vector interactions alters the energy spectrum of the model through the appearance of the new mean-field $n(\x)$ in the  Hamiltonian, and therefore the known analytical expressions for the eigenvalues of $H_0$ cannot be used. A notable exception however occurs if the density of the system is spatially homogeneous: 
In this case, as obvious from \Eq{eq:H}, the eigenvalues of  $H_0$ are just shifted by a constant amount $2G_V n$, 
and the only additional task is to determine the value of $n$ selfconsistently. 
In fact, for constant density the shifted chemical potential $\tmu$ is also constant,  
so that $\Omega_{kin}(T,\mu)$ in the presence of vector interactions is identical to $\Omega_{kin}(T,\tmu)$, evaluated for
$G_V=0$. The selfconsistent value of $\tmu$ (and thus of $n$) is then obtained from the stationarity of the thermodynamic 
potential, $\partial\Omega/\partial\tmu = 0$.

However, while for the CDW ansatz $n(\x)$ is indeed constant, this is not the case for the RKC.  
In  Ref.~\cite{CNB:2010}, in order to nevertheless make use of the known analytical expressions for the eigenvalue spectra,
the spatially modulated vector condensate in \Eq{eq:H} was approximated by its spatial average
$\bar n \equiv \langle n(z) \rangle $.  
Hence, within this ``average-density approximation'' (ADA), the above feature that the main effect of the inclusion of 
vector interactions is just a shift  $\mu \to \tmu$ in $\Omega_{kin}$ holds for all modulations, including the RKC.

In turn, this means that within the ADA the competition between different types of modulations is left substantially unaltered 
compared to the $G_V=0$ case. In particular the RKC remains favored over the CDW~ \cite{CNB:2010}. 
On the other hand, since the true density in a RKC is not constant, while it is in a CDW, one may ask to what extent
this result is just an artifact of the approximation. In fact, since the vector interaction is repulsive, it is conceivable that 
homogeneous density distributions, as in the CDW, eventually become favored over modulated ones, as in the RKC,
when the vector coupling is increased. 
In the following we will investigate this possibility.

\section{Ginzburg-Landau expansion} 
\label{sec:gl}

In the vicinity of the LP, where the homogeneous and inhomogeneous phases with broken chiral symmetry meet 
with the restored phase,  the mass function $M(\x)$ and its gradient are small, so that  
some insight into the properties of the system can be inferred via a 
Ginzburg-Landau (GL)  analysis of the thermodynamic potential. 
In this section we want to apply this technique to study the effect of vector interactions. 

To that end, following \cite{CNB:2010}, we write the shifted chemical potential as the sum of its value 
in the restored phase and  a small deviation,
$\tmu(\x) = \tmu_0^{rest} + \delta\tmu(\x)$,
and expand $\Omega$ in powers of the mass function $M(\x)$ and $\delta\tmu(\x)$ around the chirally restored solutions ($M,\tmu) = (0, \tmu_0^{rest})$. More specifically, we write $\Omega = \Omega_0 + \frac{1}{V} \int d^3x\, \Omega_{GL} $ with $\Omega_0$ the thermodynamic potential in the restored phase and 

\begin{align}
\Omega_{GL} & = \alpha_2 |M|^2 + \alpha_4 \left( |M|^4 + |\nabla M|^2 \right) 
 + \alpha_6 \left(|M|^6 + 3|\nabla M|^2 |M|^2 +
 \frac{1}{2} \left(\nabla(|M|^2) \right)^2 + \frac{1}{2} |\nabla^2 M|^2 \right) \nonumber\\
          &    + \beta_{41} (\delta\tmu)^2 + \beta_{42} |M|^2 \delta\tmu  
              + \beta_{61} (\nabla\delta\tmu)^2 + \beta_{62} (\delta\tmu)^3 + \beta_{63}  |\nabla M|^2\delta\tmu  \nonumber\\ & +  
             \beta_{64}  |M|^2 (\delta\tmu)^2 + \beta_{65}  |M|^4 \delta\tmu + \beta_{66} (\nabla^2 \delta\tmu )|M|^2 
             \; +\;  \dots \,,
             \label{eq:omegaGL1}
 \end{align} 
where the terms with the coefficients $\alpha_i(T,\mu)$ correspond to the potential for $G_V= 0$,
while the terms with the coefficients $\beta_i(T,\mu)$ appear when vector interactions are included. 
A remarkable feature of the former part is that the coefficients $\alpha_{41}$ of $|M|^4$ and $\alpha_{42}$ of
$|\nabla M|^2$ are equal ($\equiv \alpha_4$),
with the consequence that for $G_V= 0$ the LP, given by $\alpha_2 = \alpha_{42} = 0$,
coincides with the CP, which is given by $\alpha_2 = \alpha_{41} = 0$ \cite{Nickel:2009ke}. 
 
In  \Eq{eq:omegaGL1} we included terms up to order $M^6$ in order to be able to determine the favored type of modulation for the chiral condensate \cite{Nickel:2009ke,Abuki:2011pf}.
Our power counting is given by assuming that $\nabla \sim M$, as customary, while $\delta\mu$ is of order $M^2$
\cite{CNB:2010}. 
Indeed, by solving the Euler-Lagrange equation for $\delta\tmu$ 
we find to up to order $M^4$
\begin{align}
\delta\tmu = & - \frac{1}{2} R_\beta |M|^2  + \frac{1}{2 \beta_{41}}\left( -\frac{3}{4} R_\beta^2 \beta_{62} + R_\beta \beta_{64} - \beta_{65} \right) |M|^4 \nonumber\\
& - \frac{\beta_{63}}{2\beta_{41}} |\nabla M|^2 - \frac{1}{2\beta_{41}} \left( R_\beta \beta_{61} + \beta_{66} \right) \nabla^2(|M|^2) \,,
\label{eq:deltamuEL}
\end{align}
where we defined $R_\beta = \beta_{42}/\beta_{41} $. 
To leading order we thus have $\delta\tmu \sim M^2$, as claimed above.
Note that for a CDW {modulation, \Eq{eq:mzCDW},} 
all the quantities {on the right-hand side of}
\Eq{eq:deltamuEL} 
are spatially constant, so that the shifted chemical potential and therefore the density
are  constant as well.

By plugging the solution of \Eq{eq:deltamuEL} into  \Eq{eq:omegaGL1} and truncating again at ${\cal{O}}(M^6)$ we arrive at  
 \begin{align}
 \Omega_{GL} & = \alpha_2 |M|^2 + \left(\alpha_4 - \frac{1}{4} R_\beta \beta_{42} \right) |M|^4 +  \alpha_4 |\nabla M|^2 \nonumber\\
 & + \left( \alpha_6 - \frac{1}{8} R_\beta^3 \beta_{62} + \frac{1}{4} R_\beta^2 \beta_{64} - \frac{1}{2} R_\beta \beta_{65} \right) |M|^6 
 + \left(3 \alpha_6 -\frac{1}{2} R_\beta \beta_{63}  \right) |\nabla M|^2 |M|^2  \nonumber\\ 
 & + \left(\frac{1}{2}\alpha_6 + \frac{1}{4} R_\beta^2 \beta_{61} + \frac{1}{2} R_\beta \beta_{66} \right) \left(\nabla(|M|^2) \right)^2 + \frac{1}{2} \alpha_6 |\nabla^2 M|^2 \,.
         \label{eq:omegaGL2}
 \end{align}

The computation of the GL coefficients is tedious but straightforward. 
The $\alpha_i$ have already been obtained in Ref.~\cite{Nickel:2009ke} for the case $G_V=0$. 
Taking into account the shift $\mu \rightarrow \tmu_0^{rest}$ due to the vector condensates  one finds for $G_V\neq 0$
\begin{align}
\alpha_2(T,\mu) &= \frac{1}{4G_S} -\frac{N_f N_c}{8\pi^2} \sum\limits_{j=1}^3 j\Lambda^2 \ln(j\Lambda^2)
                               + \frac{N_f N_c}{4} \left( \frac{T^2}{3} + \frac{(\tmu_0^{rest})^2}{\pi^2}\right) \,,
\nonumber\\
\alpha_4(T,\mu) &= 
-\frac{N_f N_c}{8\pi^2} \left[ 1 - \int_0^\infty dp \, p^2 \, \left( \sum_j c_j \frac{1}{(p^2 + j\Lambda^2)^{3/2}} - \frac{1}{p^3}\left( n + \bar n \right)\right) \right] \,, 
\nonumber\\
\alpha_6(T,\mu) &= 
-\frac{N_f N_c}{48\pi^2} \int_0^\infty dp \, p^2 \, \left[ \sum_j c_j \frac{3}{(p^2 + j\Lambda^2)^{5/2}} 
- \frac{3}{p^5}\left( n + \bar n \right) 
\right.
\nonumber\\
&
\left. \phantom{\sum_j c_j } \hspace{30mm}
+ \frac{3}{p^4}\left( \frac{\partial n}{\partial p} + \frac{\partial \bar n}{\partial p} \right)  
 -\frac{1}{p^3}\left( \frac{\partial^2n}{\partial p^2} + \frac{\partial^2 \bar n}{\partial p^3} \right)  \right] \,,
\label{eq:glalphas246}
 \end{align} 
where we have employed Pauli-Villars (PV) regularization
with cutoff mass $\Lambda$ and coefficients $c_0=-c_3=1$, $c_1=-c_2=-3$
to regularize the diverging vacuum contributions. We also introduced
$n(p) = [\exp((p-\tmu_0^{rest})) +1]^{-1}$ and $\bar n= [\exp((p+\tmu_0^{rest})) +1]^{-1}$, corresponding to
Fermi occupation functions for 
massless quarks and antiquarks, respectively, at temperature $T$ and chemical potential $\tmu_0^{rest}$.

For the $\beta_i$ of order ${\cal{O}}(M^4)$ we obtain 
\begin{align}
\beta_{41}(T,\mu)  &= -\frac{1}{4 G_V} - \frac{N_f N_c}{4} \left( \frac{T^2}{3} + \frac{(\tmu_0^{rest})^2}{\pi^2} \right)  \,, 
\nonumber\\
\beta_{42}(T,\mu)  &=  \frac{N_f N_c}{2\pi^2} \tmu_0^{rest} \,,
\label{eq:beta4}
\end{align}
and thus 
\beq
       R_\beta = - \frac{6 G_V N_f N_c \, \tmu_0^{rest}}{3\pi^2 + G_V\, N_f N_c \left[ \pi^2T^2 + 3(\tmu_0^{rest})^2 \right]}\,,
\label{eq:Rbeta}
\eeq
which is negative for positive values of $G_V$.
Note that these coefficients, which multiply powers of $\delta\tmu$, do not have any divergent parts and therefore remain unregularized. This can easily be understood by 
observing that these terms can be obtained from the thermodynamic potential for {\it homogeneous} order parameters 
as derivatives with respect to $\tmu$. 
Hence they originate from the unregularized medium contribution to  the thermodynamic potential and therefore remain 
unregularized as well. This argument does not apply to gradient terms, which can in principle carry a regularization dependence.

From \Eq{eq:omegaGL2} we can now
see that when $G_V$ (and consequently $R_\beta$)  is nonzero, the GL coefficient multiplying $|M|^4$ acquires an additional contribution proportional to $R_\beta$ which leads to the known shift of the CP to lower temperatures \cite{Fukushima:2008b,CNB:2010}. On the other hand, the $|\nabla M|^2$ coefficient is the same as for the $G_V=0$ case and is just affected by the inclusion of vector interactions through the shift $\mu \to \tmu_0^{rest}$. As a consequence, the LP moves only along the chemical potential direction and the inhomogeneous phase is not suppressed \cite{CNB:2010}.

Let us now investigate how the inclusion of vector interactions affects the competition between different kinds of modulations close to the LP.
As already mentioned, throughout most of the inhomogeneous window the RKC at $G_V=0$ takes a sinusoidal shape. 
Indeed, performing a Fourier expansion of \Eq{eq:mzsoli},
one finds that the mass function is dominated by the lowest mode, unless the elliptic modulus is very close to $\nu=1$, 
which is only the case in the vicinity of the phase boundary to the homogeneous chirally broken phase \cite{Carignano:2017uyp}. 
In the following, we shall therefore consider a single cosine\footnote{The Fourier decomposition of  \Eq{eq:mzsoli} yields 
a sine but, of course, a cosine is equivalent, since  we are free to shift our coordinate frame without changing the free energy.}
as a prototype for real modulations with modulated density, namely we introduce
\begin{align}
 M_{COS}(z) = \Delta \cos(qz) \,.
 \label{eq:mzcos}
 \end{align}

In order to investigate the competition between the cosine with corresponding modulated density and CDW solutions close to the LP, we start from \Eq{eq:omegaGL2} and plug into it the two different Ans\"atze, \Eq{eq:mzcos} and \Eq{eq:mzCDW}. 
After minimizing with respect to $q$ in both cases, we can compute the GL free energy difference between the two different solutions. We find
\beq
\delta\Omega = \Omega_{COS} - \Omega_{CDW} = \frac{\Delta^4}{4} \left\lbrace \alpha_4 - R_\beta \left[ \frac{1}{8} \beta_{42} + \frac{\alpha_4}{\alpha_6} \left( \frac{1}{4} \beta_{63} + \beta_{66} + \frac{1}{2} R_\beta \beta_{61} \right) \right] \right\rbrace + {\cal O} (\Delta^6) \,,
\label{eq:omeGLdiff}
\eeq
For $G_V=0$ this expression reduces to  $\delta\Omega = (\alpha_4/4) \Delta^4$ \cite{Nickel:2009wj}. 
Since $\alpha_4 <0$ in the inhomogeneous phase (otherwise nonzero mass gradients would not be favored), this shows that 
in the absence of vector interactions the cosine solution has a lower free energy than the CDW \cite{Nickel:2009wj}. 

At the LP, on the other hand, $\alpha_4$ vanishes.
Very close to the LP $\delta\Omega$ is therefore dominated by the  $-R_\beta \beta_{42}$ term which is finite and positive 
for $G_V > 0$ (\cfr{}  Eqs.~(\ref{eq:beta4})  and (\ref{eq:Rbeta})). 
We thus find that already for arbitrarily small values of $G_V$ there is always a region close to the LP where the CDW 
is favored over the cosine modulation.

Away from the LP, the other terms in \Eq{eq:omeGLdiff} can play a role as well.
The missing coefficients are
\begin{align}
\beta_{61} 
   & = \frac{N_f N_c}{36\pi^2} - \frac{2}{3} \alpha_4 \,,  
   \\[1mm] 
\beta_{63} & = \frac{N_f N_c}{8\pi^2} \int_0^\infty dp\, \frac{1}{p} \left( \frac{\partial n}{\partial p} - \frac{\partial \bar{n}}{\partial p} \right)  \,, \\
\beta_{66} & = -\frac{1}{3} \beta_{63} \,,
\end{align}
where the vacuum part of $\beta_{61}$ is again PV regularized, while the two other  coefficients contain only medium contributions
and thus remain unregularized.

\begin{figure}
 \begin{center}
  \includegraphics[width=.4\tw]{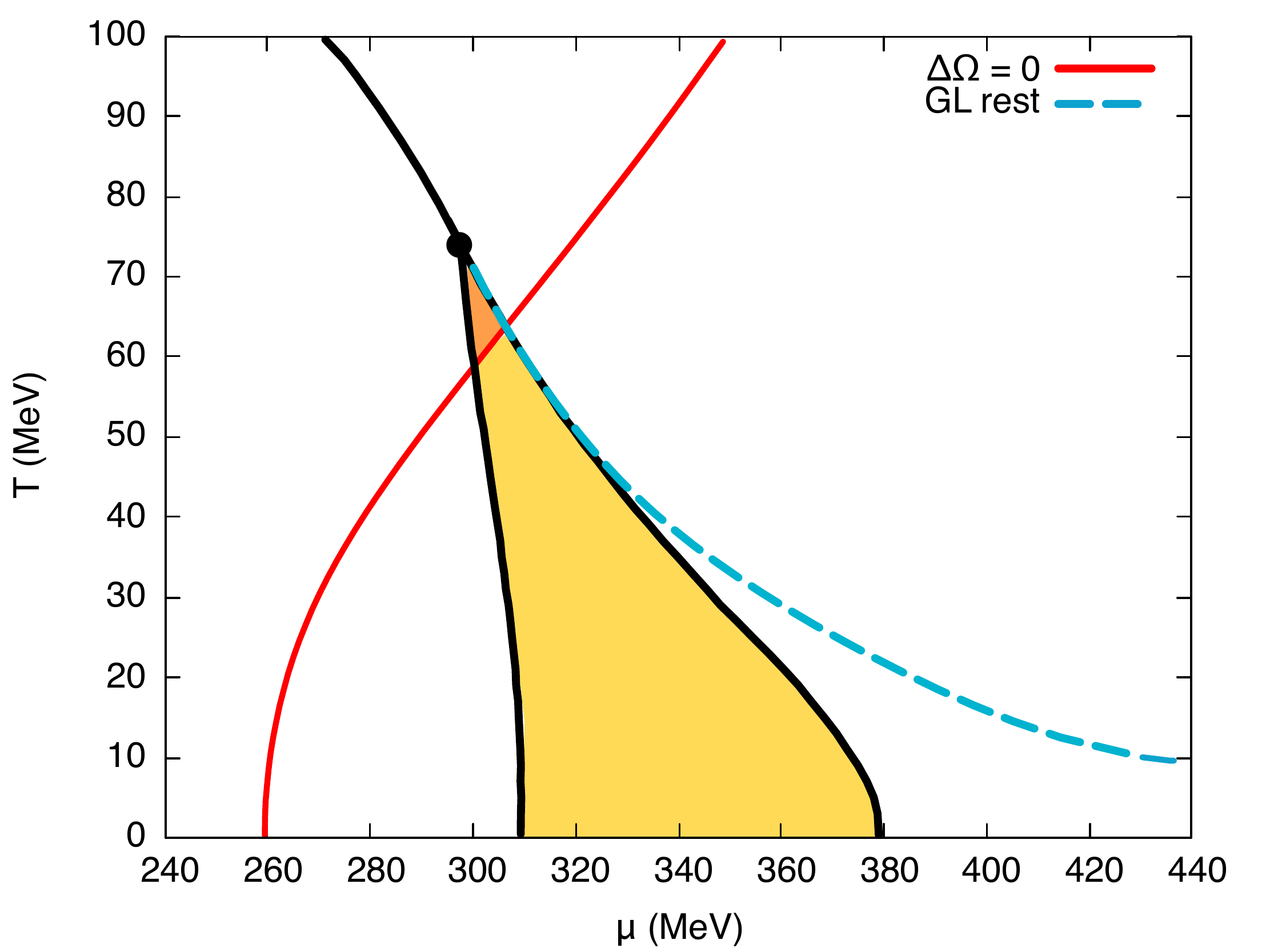}
    \includegraphics[width=.4\tw]{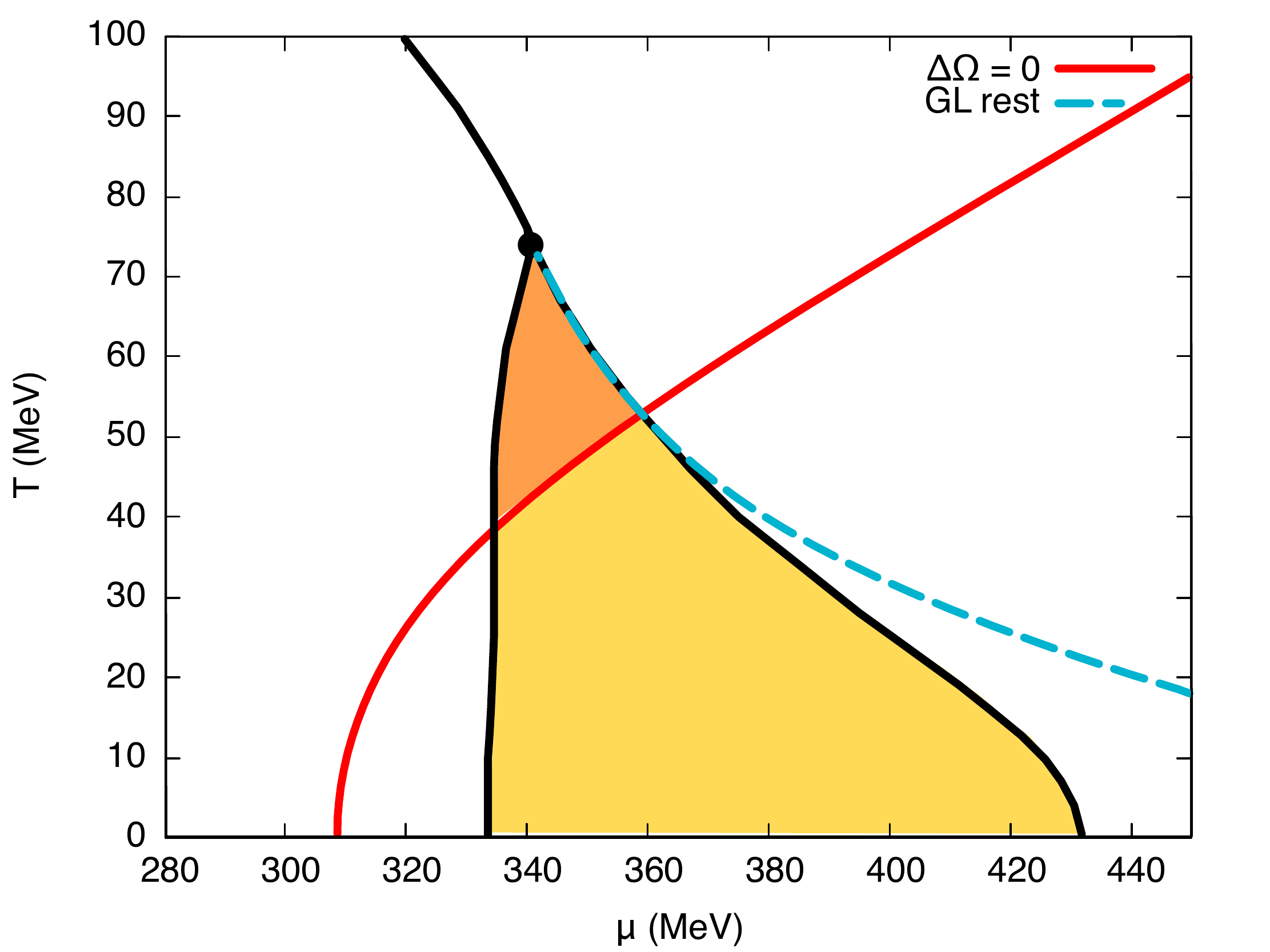}
 \end{center}
\caption{
$\delta\Omega/\Delta^4 = 0$ curve in leading-order GL approximation (red solid line)
overlayed with the phase diagram obtained in a full mean-field calculation when
allowing for CDW modulations.
 Left: results for $G_V = G_S/5$, right: results for $G_V = G_S/2$. 
In the orange-shaded region above the red line, 
the combined analysis predicts that the CDW modulation is favored over the cosine. 
In order to check the validity of the GL approximation, we also show the line 
$\alpha_4 - \sqrt{2\alpha_2\alpha_6} =0 $ 
(dashed blue), which indicates the leading-order GL result for the  phase transition from the inhomogeneous to the 
restored phase~\cite{Nickel:2009ke}. 
\label{fig:deltaomezero}}
\end{figure}

Going deeper into the inhomogeneous phase the cosine could eventually become favored again 
in particular due to the effect of the negative $\alpha_4$ term. 
Within the given GL approximation the line which separates the two regions is given by $\delta\Omega(T,\mu)/\Delta^4 = 0$,
which can be solved numerically.
The result for this curve is shown in \fig{fig:deltaomezero}. 
Here, as in all our numerical examples, we have 
chosen the value $\Lambda=757.048$ MeV for the PV regulator and a scalar coupling constant 
 $G_S=6.002/\Lambda^2$, which have been fixed by fitting the pion decay constant to $f_\pi=88$ MeV and the vacuum constituent 
quark mass to $M_{vac}=300$ MeV.
The vector coupling $G_V$, on the other hand, is treated as a free parameter of the same order as $G_S$.
In the figure we can see that, as $G_V$ increases, the region where the CDW is favored over the cosine modulation increases
as well. 

Of course, we should be careful since the GL expansion breaks down as one moves away from the LP and the order parameters or their gradients become large.
To get some idea about its validity, we also show the leading-order GL prediction for the phase boundary to the restored phase
(dashed lines) in comparison with the exact numerical result. This seems to indicate that the boundary between the CDW and 
cosine regimes could be trusted in both examples. 
Note however that near the phase boundary the order parameter is always small
and hence deviations can only be due to large gradients. 
In contrast, further away from the phase boundary the order parameter itself can be large as well. 
Therefore we refrain from making any quantitative statements in that regime.

In summary, the GL expansion suggests that, at least in proximity of the LP, the inclusion of vector interactions disfavors solutions with spatially modulated quark number density,
making the CDW the thermodynamically favored solution.

It might be possible to extend the regime of validity of the present GL
analysis further away from the Lifshitz point by following the
procedure developed in \cite{Carignano:2017meb}. A consistent
implementation of this so-called ``improved GL expansion" would require in
our case the computation of higher-order coefficients associated with the
vector mean-field, a task which we postpone to future work.  Instead, in order to
obtain  some complementary insight to our GL results away from the
LP, we attempt in the following  a full numerical diagonalization of the
Dirac Hamiltonian in momentum space.

\section{Numerical diagonalization beyond the average-density approximation} 
\label{sec:num}

In this section we attempt a full numerical diagonalization of the Dirac Hamiltonian  \Eq{eq:H} in momentum space
for given shapes of the mass function $M(\x)$ and the density $n(\x)$ or, equivalently, the shifted chemical potential $\tmu(\x)$.
Although significantly more challenging from a computational point of view, it will give us insight complementary to our GL results,
in particular about the thermodynamically favored type of modulation away from the LP.

In the following we focus again on the cosine ansatz as an approximation to the RKC and prototypical competitor to the CDW. 
In contrast to the latter, for which the density is spatially constant and therefore the ADA is exact, we expect the density of the
cosine modulation to be inhomogeneous. 
Indeed, inserting the mass function \Eq{eq:mzcos} into 
 \Eq{eq:deltamuEL} we find that in a leading-order ($\mathcal{O}(M^2)$) GL expansion  $\delta\tmu$ behaves like
$\cos^2(qz) \propto (1 + \cos(2qz))$. 
For the shifted chemical potential we therefore consider the ansatz
\begin{align}
 \tmu(z)=\tmu_0+\tmu_1\cos(2qz) \,,
 \label{eq:cos_ansatz_mu}
\end{align}
with a constant term $\tmu_0$ and a spatially modulated part with wave number $2q$ and amplitude $\tmu_1$.
Accordingly, using \Eq{eq:tmudef}, the density is given by
\begin{align}
 n(z)=
  \langle n\rangle + n_1 \cos(2qz) \,,
\end{align}
with the spatial average $\langle n\rangle = \frac{\mu-\mt_0}{2G_V}$
and an oscillating part with amplitude $n_1 = -\frac{\mt_1}{2G_V}$.
Note that this behavior is qualitatively consistent with the analytically known density profile of the RKC
at $G_V=0$~\cite{CNB:2010}, 
which also varies periodically with half of the wavelength of the mass function.

Within this setup, the Dirac Hamiltonian $H$, \Eq{eq:H}, is found to be a non-diagonal matrix in momentum space, 
since momenta which differ by $\pm q\e_z$ or $\pm 2q\e_z$ are  coupled through the oscillating components of $M$ and $\tmu$,
respectively.
The numerical determination of the eigenvalues is therefore rather involved but,  exploiting the block structure due to the 
discreteness of the sets of coupled momenta, it is  still feasible (for details see \eg{} Refs.~\cite{NB:2009,Carignano:2012sx,Buballa:2014tba}).

The resulting kinetic contribution to the thermodynamic potential, \Eq{eq:Omegakin}, is strongly divergent and needs to be regularized. 
Since sharp momentum cutoffs are known to fail when describing inhomogeneous phases~\cite{NB:2009,Nickel:2009wj},
we choose again a PV regularization scheme. 
In order to stay as close as possible to the ADA calculations of Ref.~\cite{CNB:2010},
where the PV regularization was only applied to the $G_V$-independent part $H_0$ after isolating $\tmu$
(see \Eq{eq:Hofx}), 
we now isolate the constant part  $\tmu_0$ of the shifted chemical potential
and apply the PV regularization only to the eigenvalues of the remaining part of the matrix.
In other words, we write
\beq
       (H -\mu) = \tilde H - \tmu_0 \,,
\label{eq:Htilde}
\eeq
and then regularize $ \Omega_{kin}$ as
\beq
  \Omega_{kin}
 \rightarrow
 -\frac{1}{V}\sum_{E_\lambda}  \sum_{j=0}^3 c_j \,
 T\ln\left[2\cosh\left(\frac{\tilde E_{\lambda,j}-\tmu_0}{2T}\right)\right] \,,     
\eeq
with 
\begin{align}
\tilde E_{\lambda,j}=\sgn(\tilde E_\lambda)\sqrt{\tilde E_\lambda^2+j\Lambda^2} \,,
\end{align}
where $\tilde E_\lambda$ are the eigenvalues of  $\tilde H$, and $\sgn()$ is the sign function.

For this ansatz the thermodynamically favored state for the system is then obtained by solving for the stationary points of 
the thermodynamic potential\footnote{As well-known from the homogeneous case, the stationary points
in the presence of a repulsive vector interaction correspond to maxima with respect to $\tmu$.} with respect to $\tmu_0$ and $\tmu_1$, 
\begin{align}
 \frac{\partial\Omega}{\partial \mt_0}=\frac{\partial\Omega}{\partial \mt_1}\overset{!}{=} 0 \,,
 \label{eq:gaps}
\end{align}
and minimizing it with respect to $\Delta$ and $q$. 

\begin{figure}
 \includegraphics[width=.35\tw]{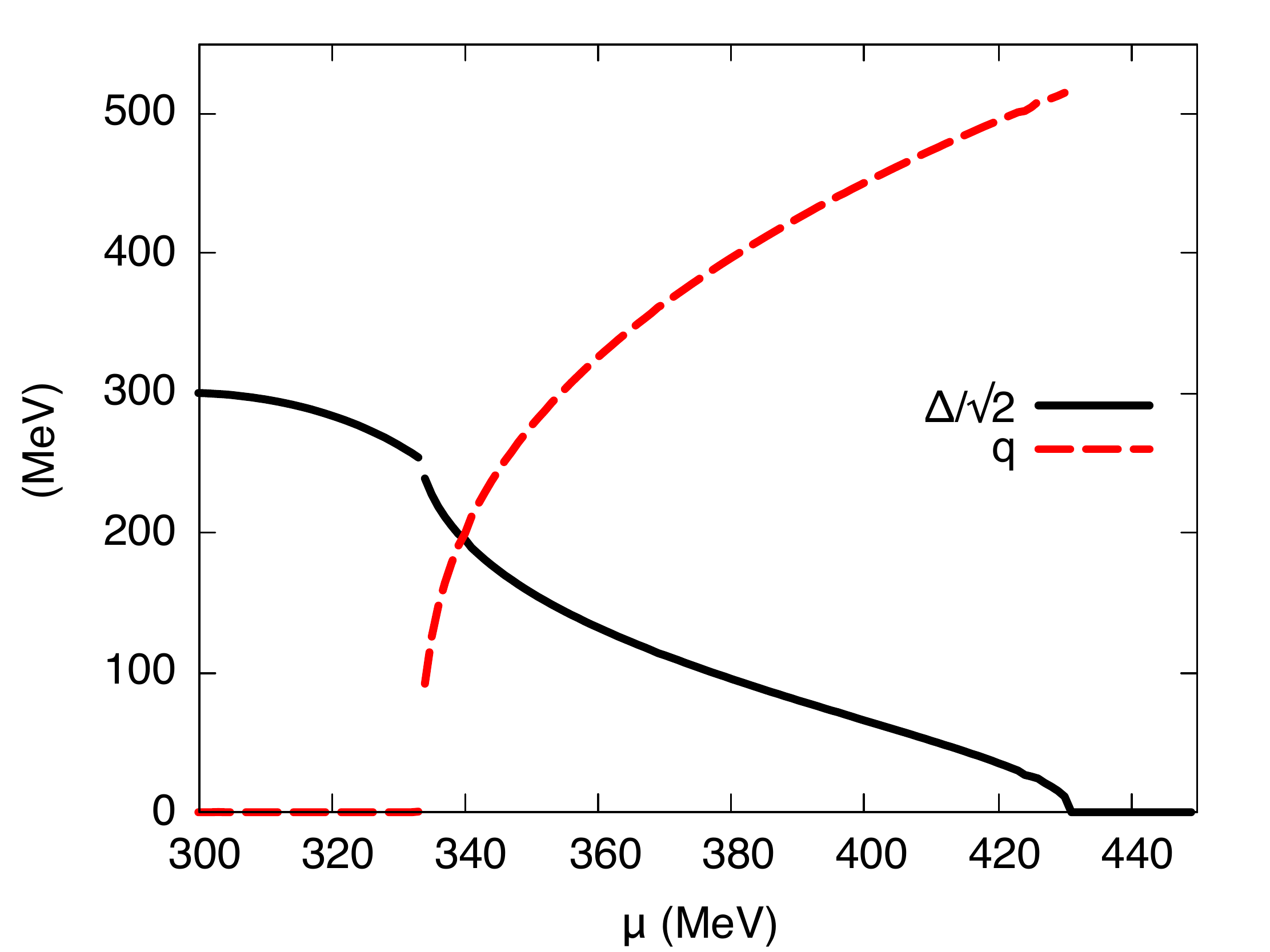}
 \includegraphics[width=.35\tw]{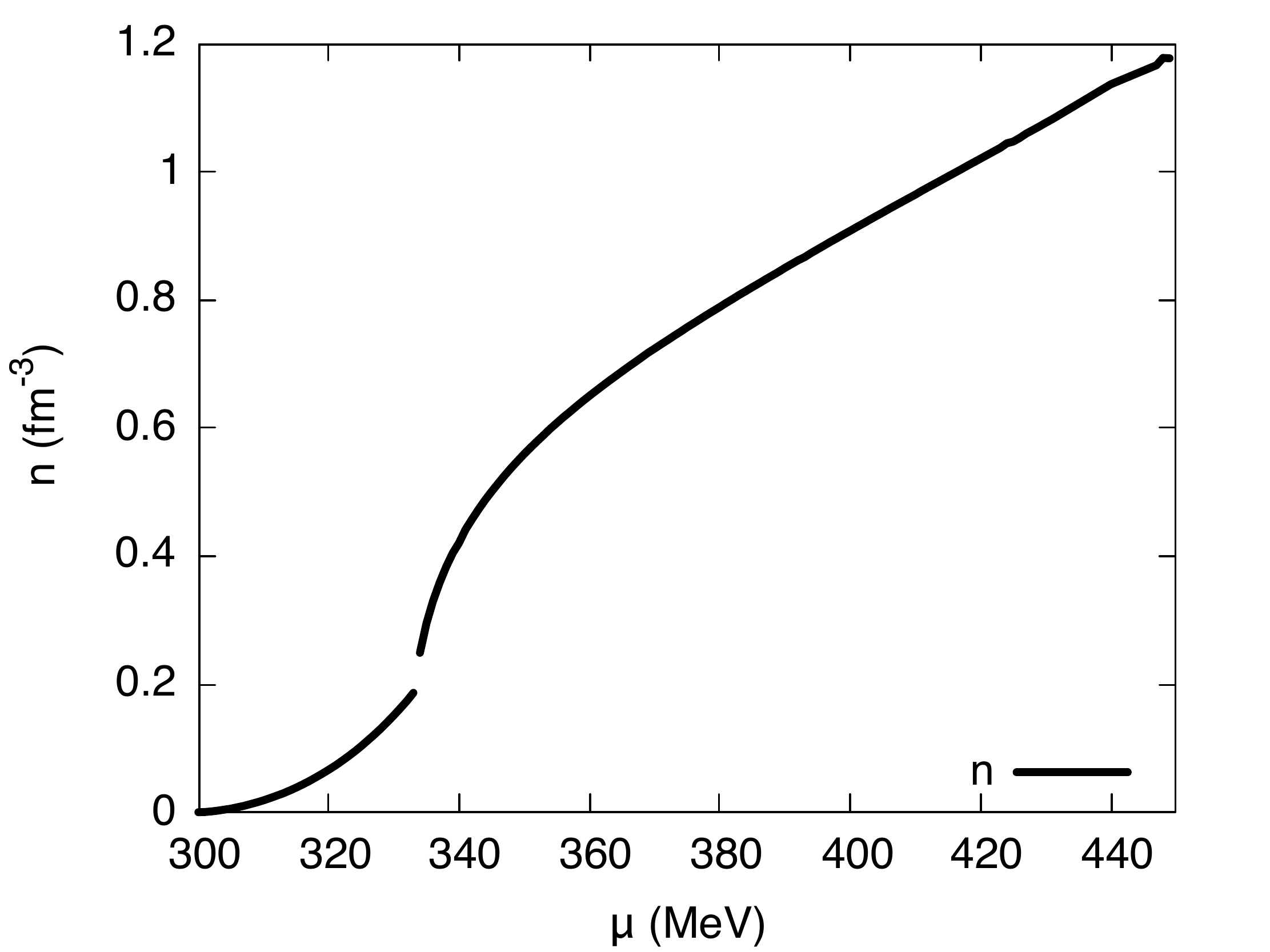}

 \includegraphics[width=.35\tw]{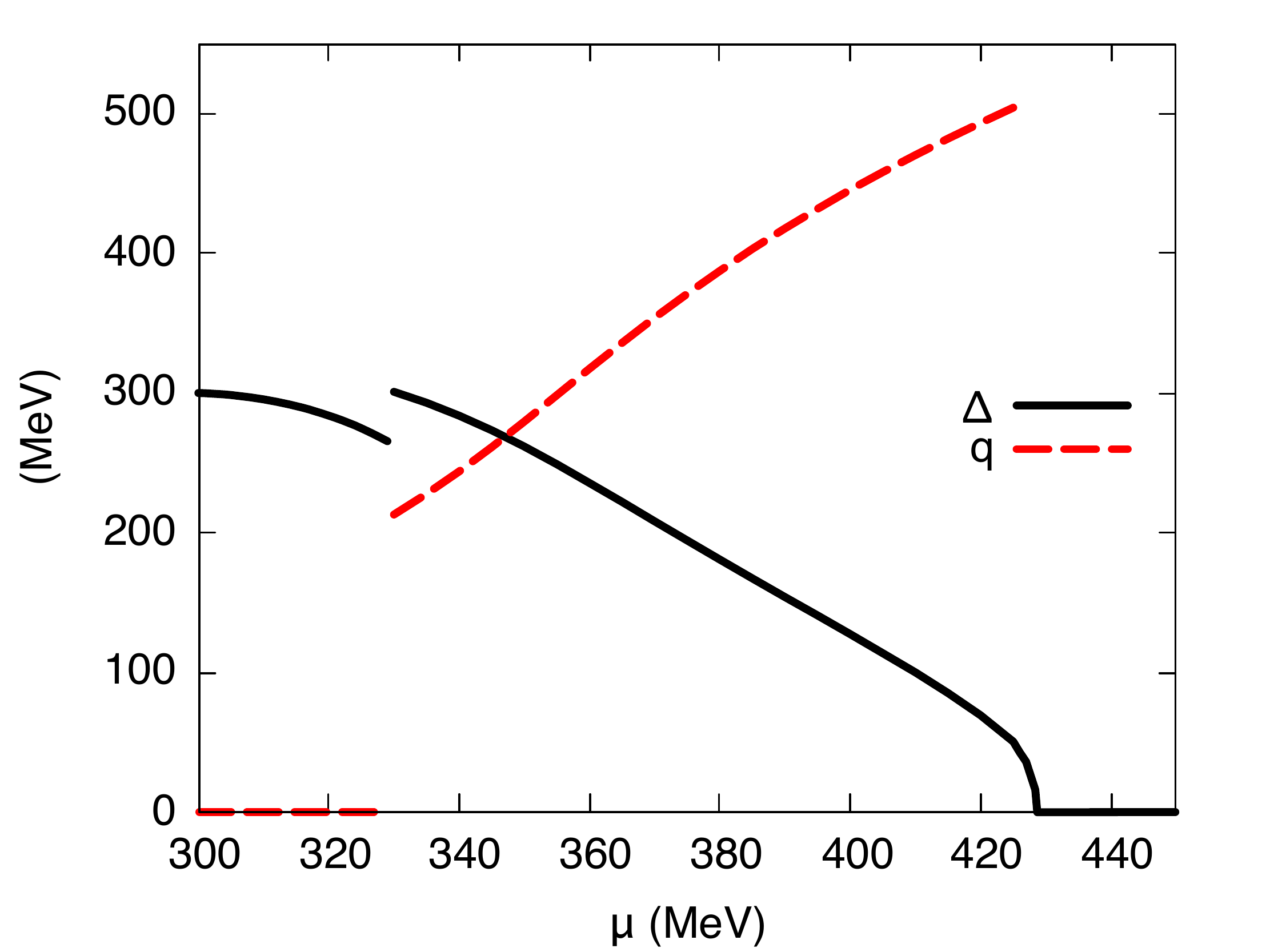}
 \includegraphics[width=.35\tw]{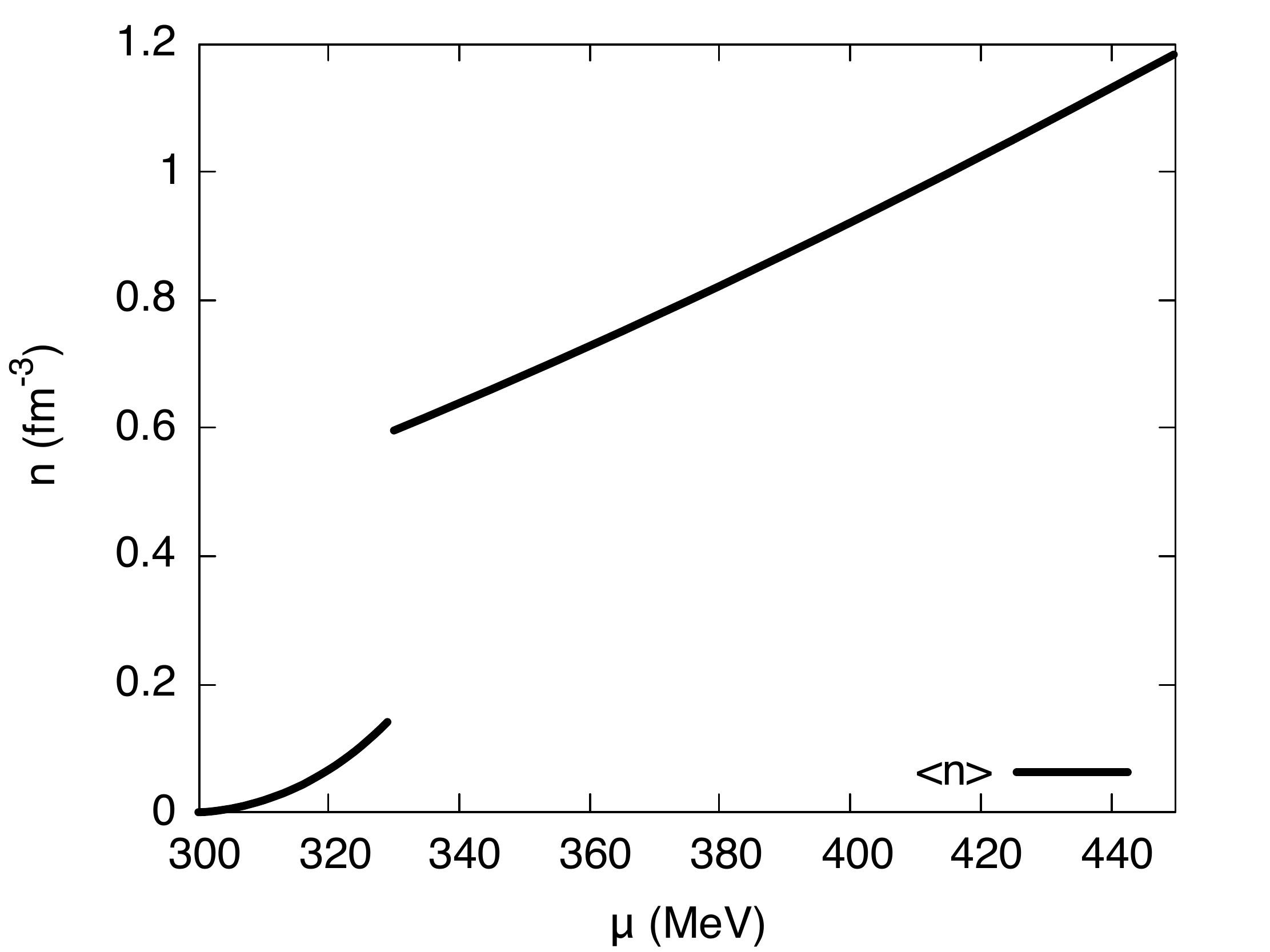}

 \includegraphics[width=.35\tw]{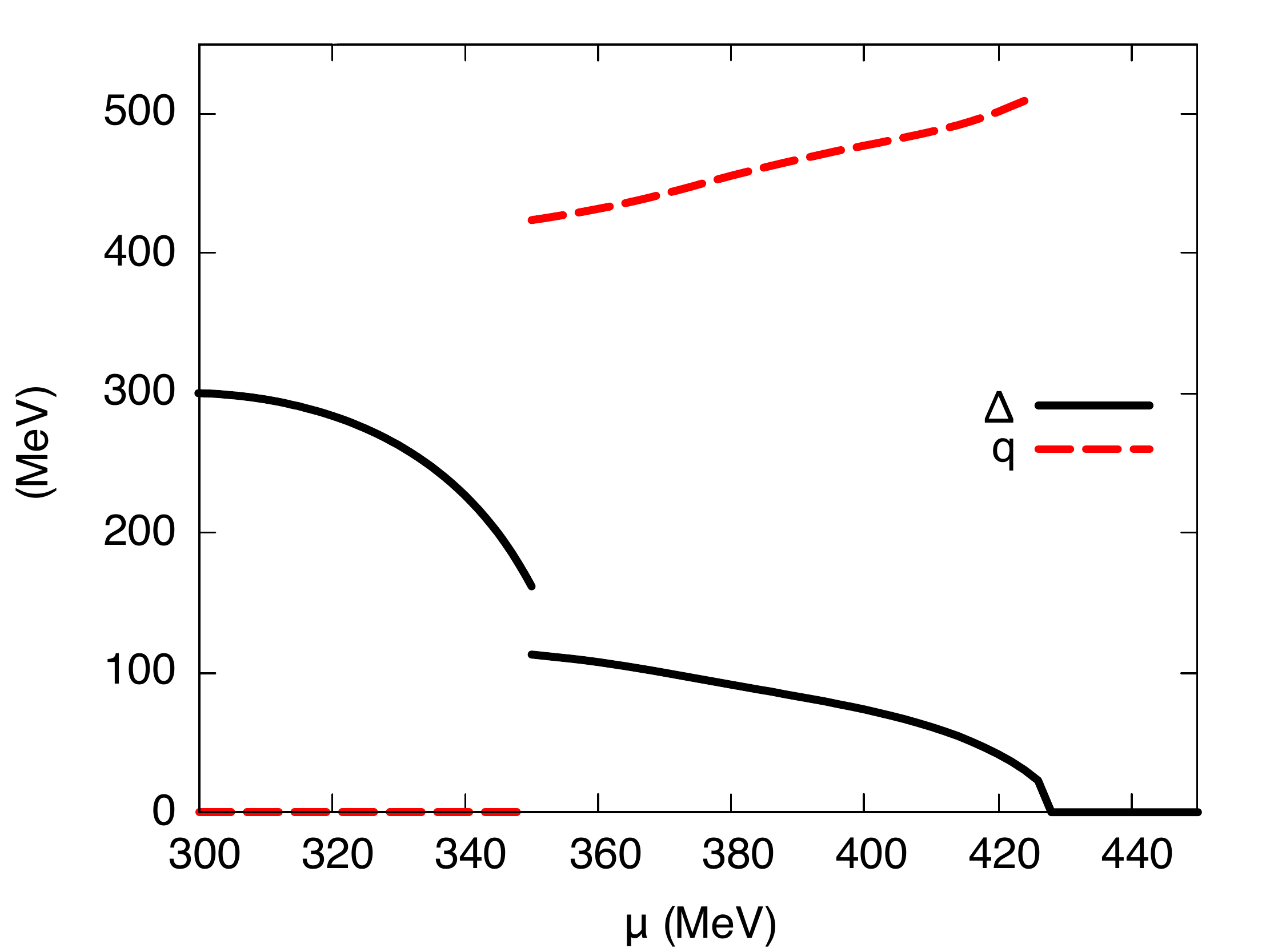}
 \includegraphics[width=.35\tw]{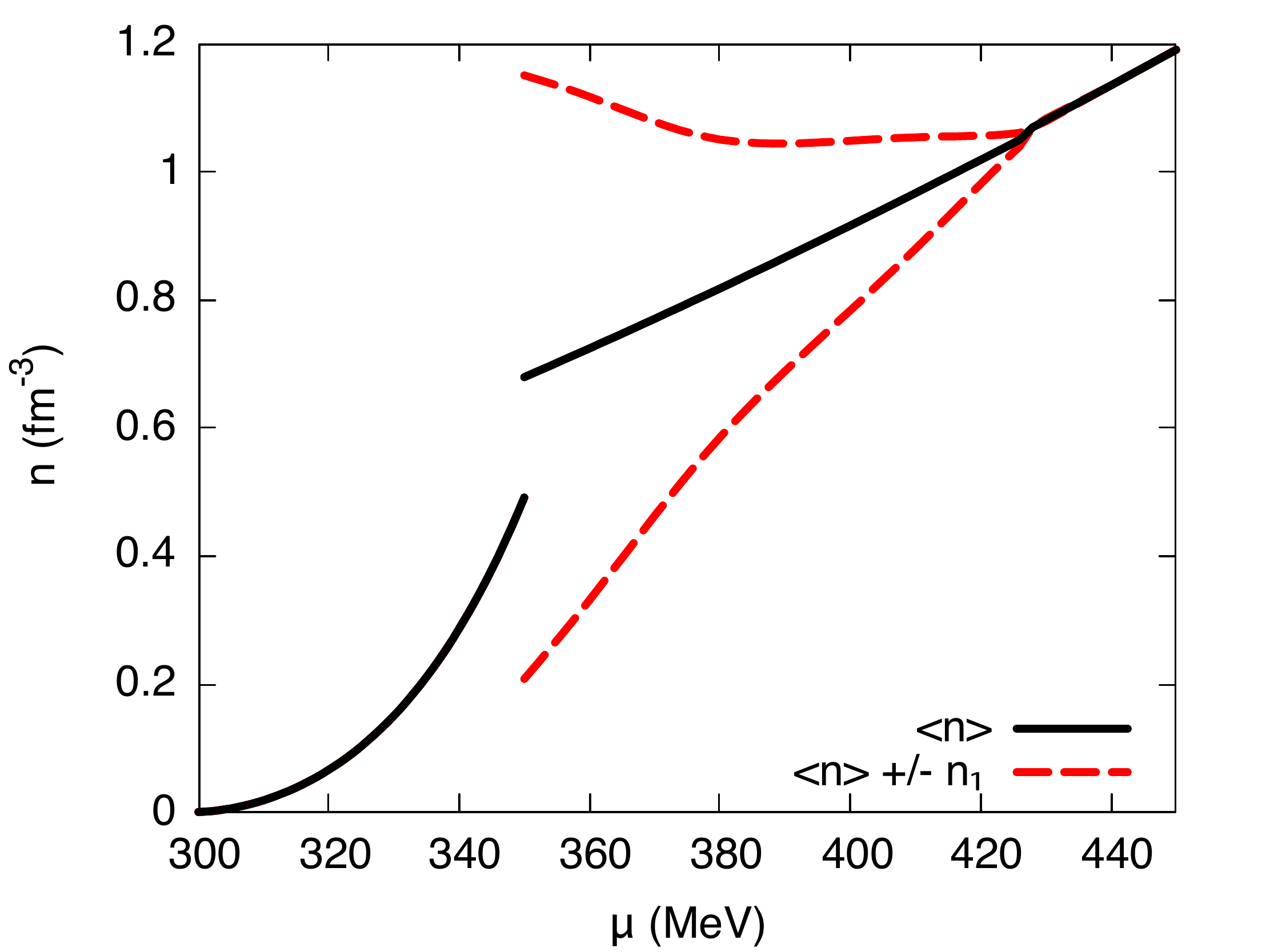}
 \caption{Left column: Comparison of order parameters at vanishing temperature as functions of the chemical potential
 for $G_V = G_S/2$ for (from top to bottom) CDW, cosine in ADA and cosine with spatially modulated density. 
 The solid lines indicate the mass amplitude, while the dashed ones represent the wave number. 
 Right column: comparison of quark number densities. Solid lines denote the average density, while the dashed lines 
 represent the maximum and minimum values of the density ( $\langle n \rangle \pm n_1$ ) in the inhomogeneous window 
 for the cosine ansatz with spatially modulated density.
 \label{fig:cosop}} 
\end{figure}

Our results for the order parameters and the density of the CDW and the sinusoidal modulation  
both within the ADA as well as with a modulated vector mean field are shown in \fig{fig:cosop}
for $G_V = G_S/2$. 
For the cosine we find that the spatial average $\ave{n}$ of the modulated vector mean field agrees quite well with the 
density obtained in ADA.
However, at the onset of the inhomogeneous phase, the amplitude $n_1$ of the density oscillations is almost as large 
as $\ave{n}$, demonstrating that the ADA is a rather poor approximation in this region.
Indeed, comparing the mass amplitudes and wave numbers of the ADA with those in the improved approach we 
find large differences, which go away only near the chiral restoration transition where the density oscillations become small.

We note that within the
 ADA the amplitude of the chiral condensate for the cosine ansatz jumps to a higher value 
 at the phase transition from the homogeneous broken
 into the inhomogeneous phase. This feature disappears when we relax the ADA and consider a modulated vector condensate.
  In this case also
 the onset of the inhomogeneous phase
  is moved towards higher chemical potentials. By comparing the critical chemical potential with that of the CDW, 
it follows immediately  
  that the CDW will become the favored modulation, at least in a certain region of the phase diagram. 

\begin{figure}
\begin{center}
\includegraphics[width=.3\tw]{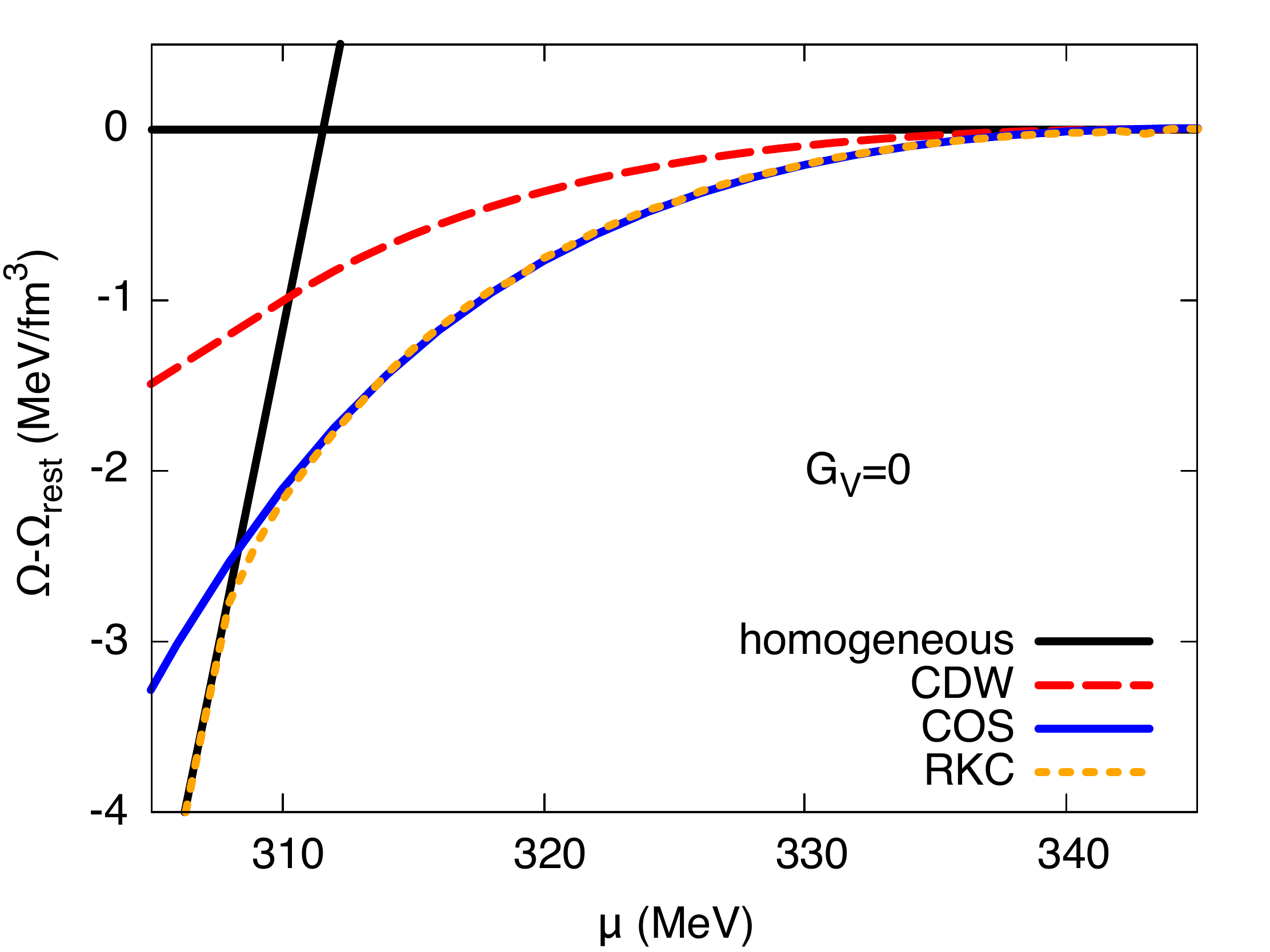}
\includegraphics[width=.3\tw]{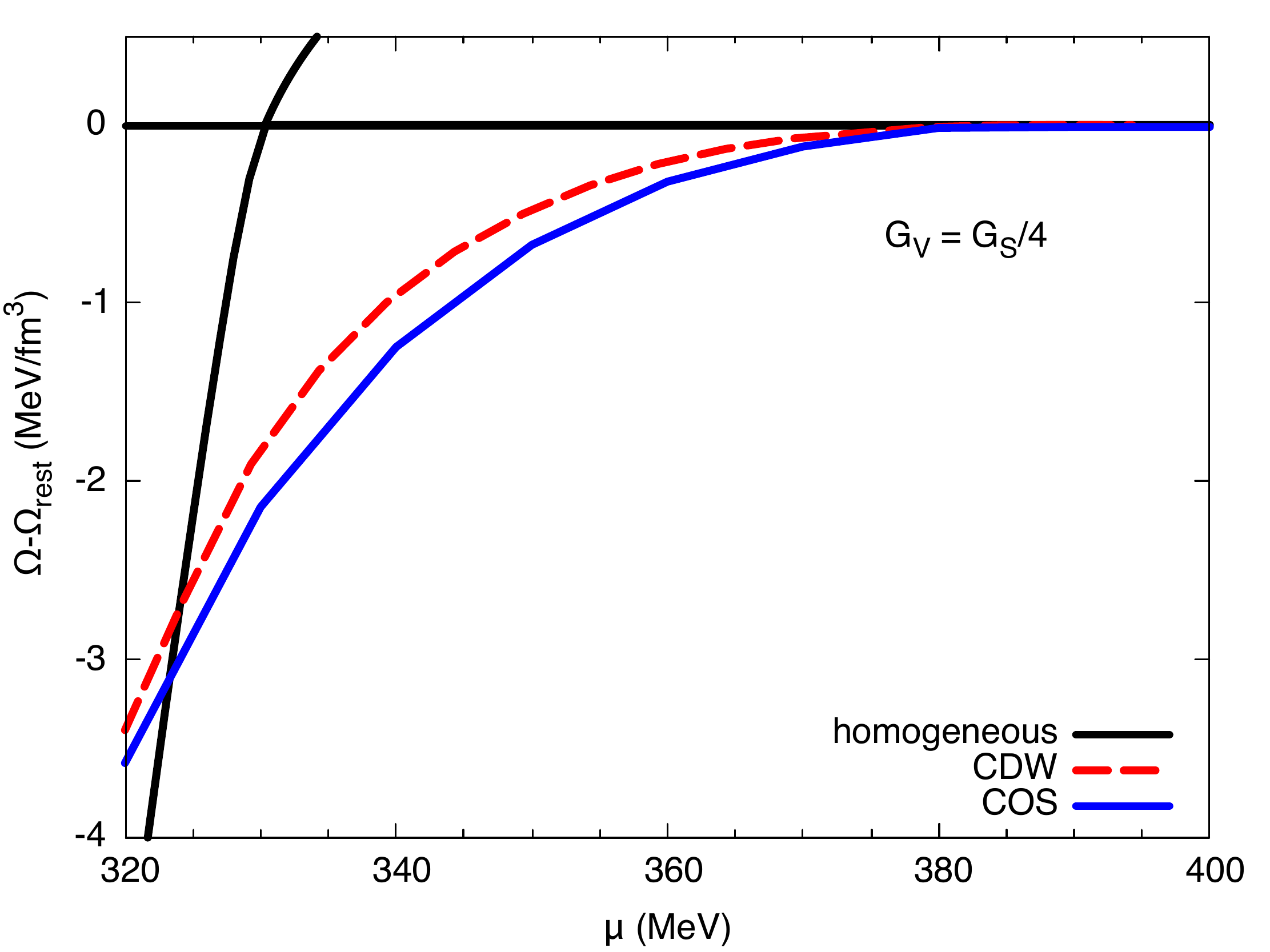}
\includegraphics[width=.3\tw]{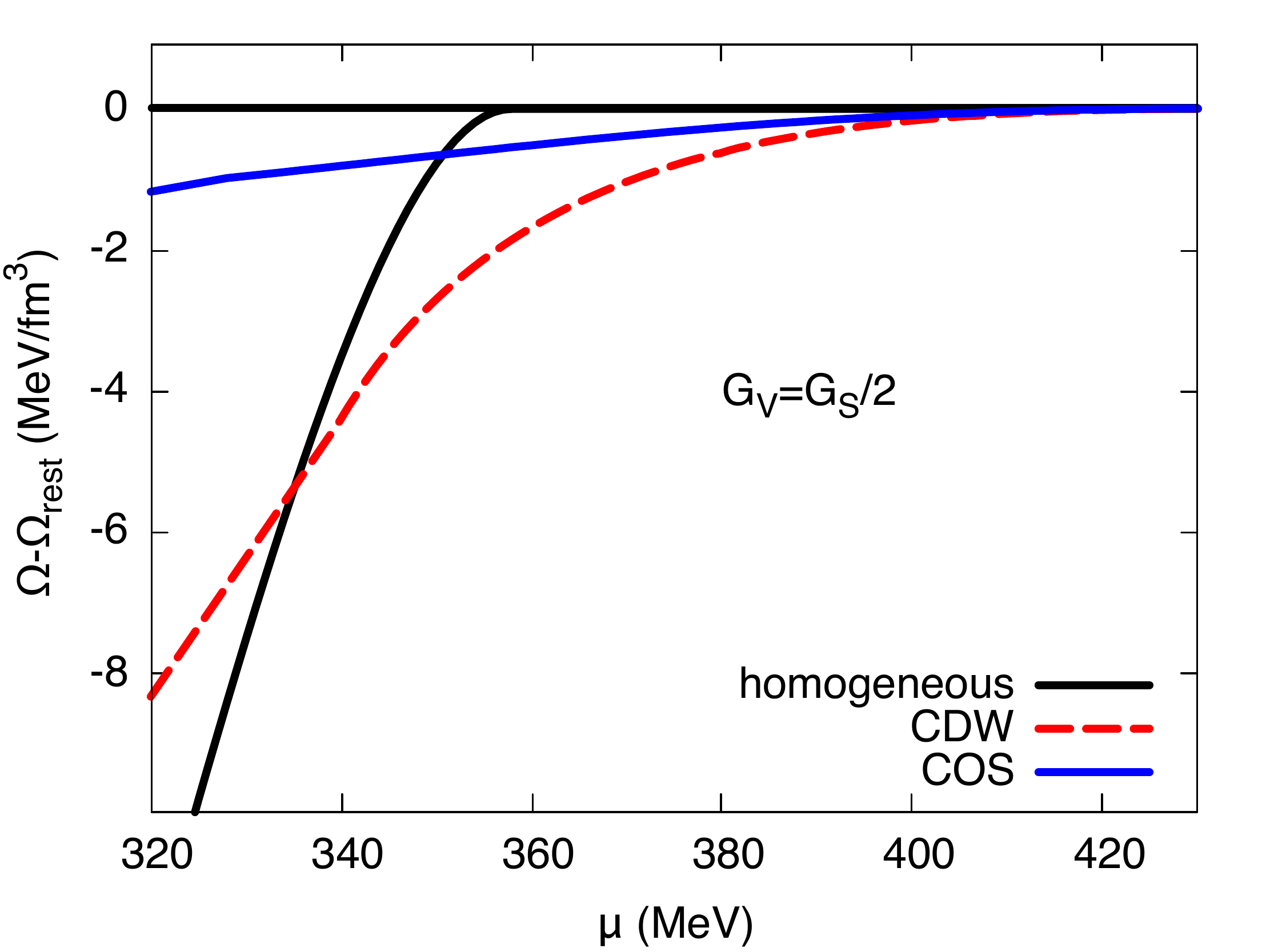}
\caption{Free energies of CDW (red dotted line) and cosine with spatially modulated 
vector condensate (blue dot-dashed) 
 at $T=0$ as functions of the chemical potential for different vector-coupling strengths, $G_V = 0, G_S/4$ and $G_S/2$, respectively. For illustrative purposes we also show in the $G_V=0$ plot the free energy associated with the RKC, in order to show that in that case this type of solution is almost degenerate with the sinusoidal one throughout the entire inhomogeneous window. 
\label{comp:energies}}
\end{center}
\end{figure}

This is confirmed by comparing the free energies associated with the different modulations, as shown 
 in  \fig{comp:energies} 
for three different values of the vector coupling. As $G_V$ increases, the CDW solution becomes favored over the cosine in an increasingly large region of the phase diagram. For intermediate values of the vector coupling we thus observe the presence of both phases within the inhomogeneous window.  
At larger $G_V$ the CDW even seems to remain favored over the cosine all the way up to the chiral restoration transition. Whether this is the case is 
hard to tell due to our insufficient numerical resolution.

%%%%%%%%%%%%%%%%%%%%%%%%%%%%%%%%%%%%%%%%%%%%%%%%%%%%%%%%%%%%%%%%%%%%%%%%%%%%%%%%%%%%%%%%
%%%%%%%%%%%%	CONCLUSION
%%%%%%%%%%%%%%%%%%%%%%%%%%%%%%%%%%%%%%%%%%%%%%%%%%%%%%%%%%%%%%%%%%%%%%%%%%%%%%%%%%%%%%%%
% \pagebreak
\section{Conclusion}
\label{sec:conclusions}

In this work, we investigated the effects of vector interactions on inhomogeneous chiral symmetry breaking within the 
NJL model, expanding on previous studies where the vector condensate was approximated by its spatial average \cite{CNB:2010}.
In the present work we allowed for a more self-consistent treatment of the problem by considering spatially modulated 
vector condensates.
Specifically, we studied a sinusoidal modulation of the order parameter in competition with a CDW, corresponding to a 
single plane wave. 
Aside from the known effect that vector interactions enlarge the inhomogeneous window in the model phase 
diagram~\cite{CNB:2010}, our main finding is that the sinusoidal modulation, which is energetically favored over the CDW
if no vector interactions are present, eventually becomes disfavored when the vector coupling $G_V$ is increased. 
In particular, a Ginzburg-Landau analysis reveals that already for arbitrarily small values of $G_V$, the CDW solution becomes favored in part of the inhomogeneous island in the model phase diagram. This result, which is expected to be valid in the proximity of the LP, is corroborated by a  numerical study at zero temperature.
This behavior can be understood by recalling that the repulsive vector interaction creates an additional energy cost if matter is not evenly spread out, so that any inhomogeneous distribution in the quark number density will become increasingly disfavored as the vector coupling becomes larger. As a consequence of this, the CDW solution, where the quark number density is homogeneous,
becomes competitive with the originally more favored sinusoidal modulation, for which the density is spatially modulated.

The window where the CDW is favored over the sinusoidal modulation increases with $G_V$. This suggests that
for sufficiently large vector couplings the CDW could be the most favored solution in the entire 
inhomogeneous phase.  
Whether this is indeed the case is however difficult to tell from our analysis since  the numerical-diagonalization method suffers
from a limited resolution, in particular near the phase boundary to the restored phase, while the GL expansion
is valid only near the LP. The improved GL expansion developed in \cite{Carignano:2017meb} might provide the tool required for this type of investigation. 

We also note that, despite the improved treatment of the vector condensate, the sinusoidal ansatz we studied is not a 
selfconsistent solution. 
While at $G_V=0$ this ansatz is a very good approximation to the RKC, which in turn is a selfconsistent solution,
we neither know the  corresponding selfconsistent solution at $G_V\neq 0$, nor whether it can still be 
well approximated by a cosine. 
It is therefore not excluded that with increasing $G_V$ the RKC could evolve continuously into a solution which persists to be 
energetically favored against both a sinusoidal shape and the CDW.
It is also possible that there is a smooth transition between the (modified) RKC and a CDW at nonzero $G_V$. 
In order to get more insight into these possibilities, it would be interesting to study more sophisticated ans\"atze with
more independent Fourier modes for both mass functions and vector condensates. 

 \section*{Acknowledgements}
 M.S.\ was supported by the GSI Helmholtzzentrum and by the Helmholtz 
Graduate School for Hadron and Ion Research. He and M.B.\ also 
acknowledge support by the Helmholtz International Center for FAIR.
 M.B.\ and S.C.\ acknowledge support by the Deutsche Forschungsgemeinschaft (DFG) through the grant
CRC-TR 211 'Strong-interaction matter under extreme conditions'.
S.C.\ also acknowledges financial support by the ``Fondazione Angelo Della Riccia".

\end{document}